\documentclass[pra,aps]{revtex4}
\usepackage{graphicx}
\usepackage{subfigure}

\begin{document}
\title{A  Robust  Quantum Random  Number  Generator  Based on  Bosonic
  Stimulation}  \author{Akshata  Shenoy   H}  \affiliation{  Dept.  of
  Electrical  and Communication  Engg., Indian  Institute  of Science,
  Bangalore,   India}  \author{S.    Omkar}  \affiliation{Poornaprajna
  Institute of Scientific  Research, Sadashivnagar, Bengaluru- 560080,
  India.}     \author{R.     Srikanth}   \email{srik@poornaprajna.org}
\affiliation{Poornaprajna    Institute    of   Scientific    Research,
  Sadashivnagar,   Bengaluru-   560080,  India.}    \affiliation{Raman
  Research   Institute,  Sadashivnagar,  Bengaluru-   560060,  India.}
\author{T.     Srinivas}   \affiliation{Dept.   of    Electrical   and
  Communication Engg., Indian Institute of Science, Bangalore, India.}

\begin{abstract} 
We  propose  a  method  to  realize a  robust  quantum  random  number
generator  based on bosonic  stimulation. A  particular implementation
that   employs  weak  coherent   pulses  and   conventional  avalanche
photo-diode detectors (APDs) is discussed.
\end{abstract}

\maketitle
  
\section{Introduction}

Random numbers  are crucial for  various tasks, among  them generating
cryptographic  secret keys,  authentication,  Monte-Carlo simulations,
digital   signatures,  statistical   sampling,  etc.    Random  number
generators  can be  classified  into two  types: pseudo-random  number
generators  (PRNG)  \cite{prng}  and  true  random  number  generators
(TRNG).   A  PRNG is  an  algorithm,  computational  or physical,  for
generating a  sequence of numbers that approximates  the properties of
random numbers.  A physical or  hardware version is typically based on
stochastic  noise or chaotic  dynamics in  a suitable  physical system
\cite{strng}.    Computational  PRNGs   are  based   on  computational
algorithms   that  generate   sequences  of   numbers  of   very  long
periodicity,   making  them   look  like   true  random   numbers  for
sufficiently short  sequences.  Careful observation  over long periods
will  in  principle  reveal  some  kind  of  pattern  or  correlation,
suggestive of non-randomness.

As  far as  is  known  today, the  inherent  indeterminism 
or fluctuations in  quantum
phenomena  is  the  only  source  of  true  randomness,  an  essential
ingredient in quantum cryptography \cite{gis}.  Various proposed 
underlying physical processes for
quantum  random number  generators  (QRNGs) include:
quantum measurement of single  photons
  \cite{jen,dynes}, an entangled  system \cite{owens}, coherent states
  \cite{ren,2ren} or vacuum states \cite{gabriel};
phase noise
  \cite{luo}, spin noise \cite{kats}, or radioactive decay or photonic
  emission \cite{madhuvit}. 

In  this work,  we propose  a novel  method of  QRNG that  is  a quite
different  indeterministic  paradigm  from  the above  two.   It  uses
\textit{bosonic stimulation} to  randomly amplify weak coherent pulses
to intense pulses that can  be easily detected by a conventional APDs.
Bosons (integer-spin quantum particles) obey Bose-Einstein statistics,
which entails that the transition  probability of a boson into a given
final state in enhanced by the presence of identical particles in that
state.   If there  are $N$  particles in  a given  quantum  state, the
probability that an incoming boson  makes a transition into that state
is proportional  to $N+1$. This effect is  called bosonic stimulation,
and is  responsible for coherent  matter wave amplification  in atomic
lasers, as  well as  the sustenance  of a particular  mode in  a laser
cavity, whereby  the presence of  photons in a particular  lasing mode
stimulates the emission of more photons into that mode.  It provides a
new way to  combine quantum indeterminism with the  tendency of bosons
to   congregate  indistinguishably.    We   may  call   this  QRNG   a
\textit{random bosonic stimulator}. Perhaps the practical merit of our
proposed QRNG, apart from its harnessing a novel version of quantum
indeterminism, is that it simplifies the detection module to the point
where it may be accessible to an advanced undergraduate laboratory.

\section{Bosonic stimulation as a realization of the P\'olya Urn problem}

\begin{figure}
\includegraphics[width=12cm]{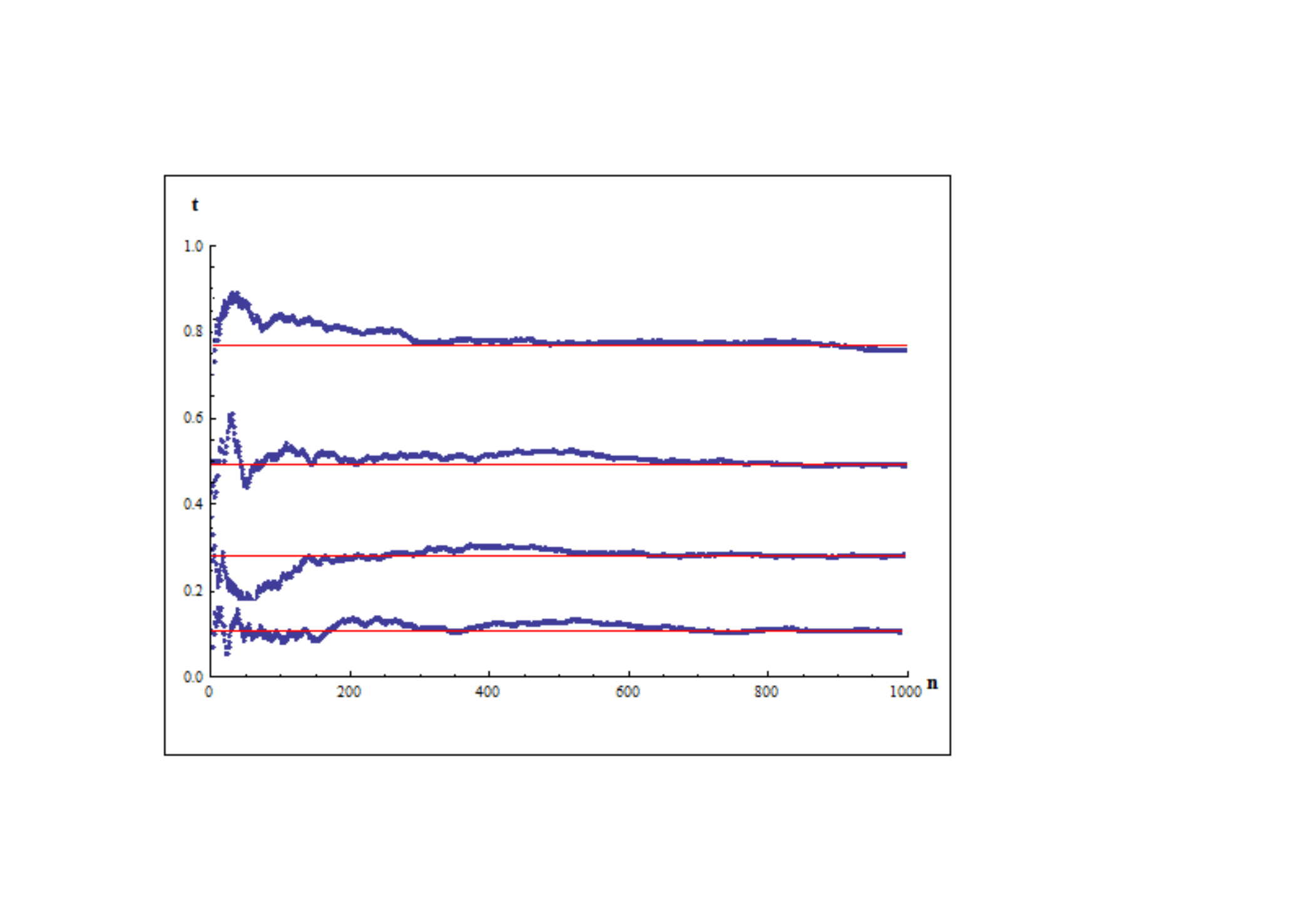}
\vspace{-1.0cm}
\caption{The relative population of the two ball populations of
the P\'olya urn, for 4 different runs 
with the same initial condition $b(0) = r(0) = 3$.}
\label{fig:graph}
\end{figure}

Consider an  urn with  $b$ blue balls  and $r$  red balls.  A  ball is
picked at random and replaced with  $c$ balls of the same color or $d$
balls of different color.  The addition of same color balls results in
positive  feedback whereas that  of different  color balls  results in
negative feedback.  During a  run of trials, the fractional population
of each  urn may initially fluctuate randomly,  but eventually settles
down  to a randomly  selected \textit{limiting  value} $t$,  giving an
instance of symmetry breaking (Figure \ref{fig:graph}).

Let  the initial  population of  two  states labelled  as ``blue"  and
``red"  be $b(0)$  and $r(0)$,  respectively.  As  the  incoming balls
start populating  the two states, the subsequent  growth in population
of the modes exhibits P\'olya urn behaviour.  The probabilistic law of
evolution of the  fractional population at $i$th instance  is given by
bosonic stimulation to be
\begin{widetext}
\begin{equation}
(b(i), r(i))\longrightarrow \left\{ \begin{array}{ll} (b(i)+1, r(i)) &
    \mbox{with probability $\frac{b(i)+1}{b(i) + r(i) + 2}$}\\ (b (i),
    r (i)+1) & \mbox{with probability  $\frac{r (i)+1}{b (i) + r (i) +
        2}$}.
\end{array} \right.
\label{eq:bostimu}
\end{equation}
\end{widetext}

The limiting  value $t \equiv  b(i)/(b(i)+r(i))$ for large  $i$ itself
varies randomly  in the range $[0,1]$  from run to run,  having a beta
distribution  
\begin{equation}
f (t;\beta,\rho )= \frac{1}{B(\beta,\rho)}t^{\beta-1}(1-t)^{\rho-1},
\label{eq:beta}
\end{equation}
where   $B$    is   the    beta   function   that    normalizes   $f$,
$\beta=b^\prime/c$, $\rho=r^\prime/c$,  $b^\prime \equiv b(0)  + s_b$,
$r^\prime \equiv r(0)+s_r$. For bosonic stimulation, the shifts $s_b =
s_r =  1$.  If  the seeding  is symmetric, then  so is  the asymptotic
distribution of the limiting values, thereby restoring symmetry, as it
should be \cite{KS}.

Depending on  the number of  trial runs, the  final state can  have an
arbitrarily large number of bosons. Depending on whether $t>0.5$ (blue
dominates) or $t<0.5$ (red dominates), one generates a random bit 0 or
1. This  can serve as  the basis of  generating random bits at  a rate
determined  by the  frequency with  which  each run  can be  repeated.
Thus,  the phenomenon  of bosonic  stimulation acts  as  a macroscopic
QRNG.

\section{Practical realization}

\begin{widetext}
\begin{figure}
\includegraphics[width=10cm]{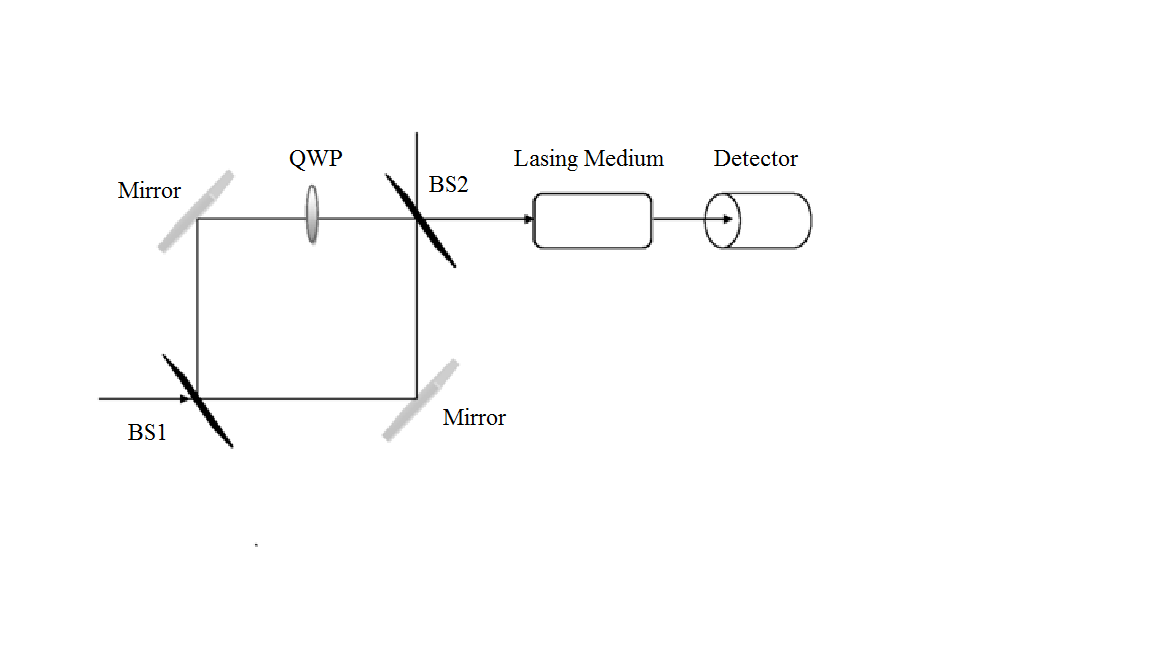}
\vspace{-2.0cm}
\caption{At the  input port, a weak  coherent pulse $|\alpha,H\rangle$
  with  horizontal polarization and  average photon  number $|\alpha|$
  enters the experiment.  The upper  arm branch is rotated to vertical
  polarization  by  the quarter-wave  plate  QWP.   The product  state
  $|\frac{\alpha}{4},V\rangle|\frac{\alpha}{4},H\rangle$   enters  the
  lasing medium,  where the $V$  and $H$ modes participate  in bosonic
  stimulation. In the symmetric case, the detector complex generates a
  0 or 1  bit depending on whether $I_H > I_V$  or the converse, where
  $I_H,  I_V$   are  intensities   of  the  horizontal   and  vertical
  polarization output components.}
\label{fig:rng}
\end{figure}
\end{widetext}

A concrete idea for realizing a  random bosonic stimulator is to use a
lasing  medium  that supports  two  radiation  modes,  for example  by
vertical and horizontal polarization of the same frequency \cite{K}. A
scheme of  the proposed experiment  is given in  Figure \ref{fig:rng}.
Two equal  intensity, highly attenuated  modes of coherent  states are
input into  a lasing medium.
To  ensure that  the two inputs are synchronized and of equal intensity,  
a calibrated Mach-Zehnder set-up is used with  an attentuated
coherent laser pulse fed
into one of its input ports. This results in an ouput consisting of
two (unentangled) coherent pulses with half the intensity.
A quarter wave plate in one of the arms ensures that the polarization 
in one arm rotated to be 90$^\circ$
with respect to the other.

Each mode  in a pulse corresponds to  a ball color in  the P\'olya urn
problem.  Because  of bosonic  stimulation, the output  intensity will
randomly favor vertical or horizontal polarization.  Let $I_H$ ($I_V$)
denote  the  intensity  of  the  outcoming  light  in  the  horizontal
(vertical) polarization mode.

The incoming photon emitted as  a result of the coherent de-excitation
of the  atoms in the medium is  assumed to be equally  coupled to both
modes. Suppose the  two modes start in the  state $|n,m\rangle$, where
the first  register corresponds to the  `blue' mode and  the second to
the  `red' mode.   Further  let the  series  of atoms in  the
population-inverted   state   be   in   the  initial   excited   state
$|e,e,\cdots\rangle$. By giving up a  photon into the blue or red mode
(which could  be angular momentum states),  the atom is  
left in the state $|b\rangle$ or $|r\rangle$, assumed to be
mutually orthogonal.  The joint system of the modes and atoms  
evolves in  a manner analogous to a quantum walk, given by:
\begin{widetext}
\begin{eqnarray}
|n,m\rangle|e,e,\cdots\rangle &\rightarrow& 
\frac{1}{\sqrt{n+m+2}}\left( \sqrt{n+1}|n+1,m\rangle|b\rangle +
   \sqrt{m+1}|n,m+1\rangle|r\rangle\right)|e\cdots\rangle \nonumber \\
 &\rightarrow& \frac{1}{\sqrt{(n+m+2)(n+m+3)}}\left( 
   \sqrt{(n+1)(n+2)}|n+2,m\rangle|b,b\rangle +
   \sqrt{(n+1)(m+1)}|n+1,m+1\rangle \times \right. \nonumber \\ && 
   \left.
   (|b,r\rangle + |r,b\rangle) +
    \sqrt{(m+1)(m+2)}|n,m+2\rangle|r,r\rangle\right)|\cdots\rangle,
\label{eq:nieuw}
\end{eqnarray}
\end{widetext}
and  so on.   Each term  in the  superposition is  rendered incoherent
because it is  entangled with a distinct state of  atoms, and thus the
probability for  scattering into a  given urn state  is quantitatively
the same as  the classical P\'olya urn situation.

Because of the
entanglement with the polarization degrees of freedom of the atoms
(Eq. (\ref{eq:nieuw})), the
state of the atoms bears an imprint of the final outcome of laser light.
However, as the laser atoms are not individually accessible, the random
bit generated is practically unique. From the view point of Monte-Carlo
simulations, etc., only the randomness from the laser light read-out 
will be used.
From a cryptographic perspective, the laser system will remain 
physically well within the encoder unit, preventing its access to a
malevolent eavesdropper. Assuming that any possible information
leakage through side-channels (like heat radiations from the laser) are
reasonably blocked out to the outside world, the assumption of practical
uniqueness of the generated randomness applies here, too.

A  random  bit  $x$  is generated  by  the  detector  module,
depending  on which  mode  dominates, for  example,  according to  the
recipe:
\begin{eqnarray}
I_H > I_V  \Longrightarrow x=0, \nonumber \\
I_H < I_V  \Longrightarrow x=1.
\label{eq:rbit}
\end{eqnarray}

The mean and variance of the distribution $f(t;\beta,\rho)$ are given,
respectively, by
\begin{eqnarray}
\mu   &=&   \frac{b(0)}{b(0)+r(0)}; \nonumber \\
\Delta^2   &=&
\frac{b(0)r(0)}{(b(0)+r(0))^2(b(0)+r(0)+1)}.
\end{eqnarray} 
The value $t$  obtained will tend to peak towards  the mean, with ever
lower variance  if one or both  of the initial  populations are large.
If  the  instrument function  of  the  detector is  
denoted by a normal distribution $e$ with FWHM $s$,  the observed
distribution is the convolution $p(x) = \int f(t;\beta,\rho)e(x-t)dt$.
It is
important that we use sufficiently weak pulses obtained by attenuating
coherent light sources, so that $s \ll \Delta$.  This ensures that the
quantum randomness  dominates over stochastic noise  in the detector's
reading.  The  distribution $f(t;\beta=1,\rho=1)$ is
uniform over $[0,1]$,  and thus the
performance  of the  random bosonic  stimulator is  least  affected by
detector  tolerance.   This  means  that the  two
inputs should ideally be vacuum modes.

A challenge will be to ensure  that the beam splitters are truly 50-50
and  the coupling  of the  excited atoms  is equal  to the  two modes.
However,  even  if  this  is  not so,  Eq.   (\ref{eq:beta})  and  the
prescription (\ref{eq:rbit})  can be  suitably generalized to  yield a
random bit of  uniform distribution.  Suppose a photon  couples to the
$H$  mode stronger  than to  the  $V$ mode  (because of  an atomic  or
beam-splitter  feature) by  a  factor $(1+  \epsilon)$,  then the  new
distribution can be shown to  be Eq.  (\ref{eq:beta}), but with $\beta
\rightarrow \beta^\prime  = \beta(1+\epsilon)$.  Let  $t_{1/2}$ be the
median    of    the    distribution    $f$,    defined    such    that
$\int_{t=0}^{t_{1/2}}  f(t; \beta^\prime,\rho)  =  \frac{1}{2}$, where
$f$  is the  beta distribution  (\ref{eq:beta}). Then,  we  obtain our
uniformly random bit by replacing prescription (\ref{eq:rbit}) by:
\begin{eqnarray}
t < t_{1/2} \Longrightarrow x=0 \nonumber \\
t \ge t_{1/2} \Longrightarrow x=1
\label{eq:rrbit}
\end{eqnarray}
To be precise, the above numbers  assume that the input modes are pure
number states.

More realistcally, taking into  account that they are coherent states,
we must replace Eq. (\ref{eq:beta}) by 
\begin{equation}
f^\prime           (t;\lambda)          =          \sum_{\beta,\rho}
\frac{1}{B(\beta,\rho)}t^{\beta-1}(1-t)^{\rho-1}P(\lambda,\beta)
P(\lambda,\rho),
\end{equation} 
where $P(\lambda,x)$  is the Poisson distribution of  $x$ with mean
$\lambda$.  Furthermore, in  practice we may have to  let $\lambda$ to
range over an interval  because of practical difficulties of producing
the same  exact degree of attentuation  on each run.  It  can be shown
that this added complication does not affect our main results.

For sufficiently low noise in each run, two bits may also be generated
per run according to the recipe:
\begin{eqnarray}
t < t_{1/4} \Longrightarrow x=00 \nonumber \\
t_{1/4} \le t < t_{1/2} \Longrightarrow x=01 \nonumber \\
t_{1/2} \le t < t_{3/4} \Longrightarrow x=10 \nonumber \\
t \ge t_{3/4} \Longrightarrow x=11,
\label{eq:2rbit}
\end{eqnarray}
where   $t_\xi$   is    defined   such   that   $\int_0^{t_\xi}   f(t;
\beta^\prime,\rho) =  \xi$. More generally, to generate  $n$ bits, the
noise level  should be lower than  $2^{-n}$.

\section{Discussion and Conclusions}

We have proposed a  novel QRNG principle, based on bosonic
stimulation, in which, while the state preparation procedure presents
experimental challenges, the detection and read-out parts are
easier to implement. In  an actual experiment, it is  possible that systematic
experimental biases might introduce  correlations into the sequence of
bits produced, thereby degrading the randomness.  Statistical analyses
like the Diehard tests and National Institute of Standards and Testing
(NIST) suite  of tests for randomness  have to be carried  out to know
the quality of randomness and improve upon it.

Some implementational details are worth noting.
Since the required light is not for communication to a geographically 
distant station, any laser system that can be conveniently tuned to 
operator at 800 nm, where good detectors (APDs) are available, is 
suitable for our purpose. A good candidate is femtosecond 
Titanium-Sapphire lasers, for which pulse repetition rates upto a
few gigahertz can be obtained. 

Further, the potential problem posed by the cavity's unequal coupling 
to the two modes can taken care of by ``biasing'' the comparator 
(as clarified above).  Care must be taken to choosing the 
operational/differential amplifier (op amp) IC designed to work as 
comparator, so that their frequency specifications are appropriate 
for the pulse rate of the laser. Even with the low resolution provided
by two-bit comparators, for a Titanium-Sapphire laser, mentioned
above, this yields a random bit-rate of the order of $10^9$ bits per second,
which is comparable to bit-rates in state-of-the-art QRNGs.

As another possible realization of bosonic stimulation based QRNG, 
one may consider photonic band gap materials, results pertaining
to which are presented elsewhere.

The  present   method  of  generating  randomness  is   based  on  the
essentially quantum feature  of being able to turn  a distinct bosonic
particle into an indistinguishable  part of a collective object, which
is  the mode or  a condensate.   The notion  of identity  in classical
logic and philosophy is as such incapable of handling this situation,
requiring recourse  to concepts like quasi-set  theory \cite{nava}. We
think that, as  in the case of the  application of quantum nonlocality
to  cryptography  \cite{ekert}, our  method  of generating  randomness
based  on  bosonic  indistinguishability,  can perhaps  be  termed  as
another instance of \textit{applied philosophy}! 

Here  we   have  presumed  the  axiomatic   indeterminism  of  quantum
mechanics. This  assumption would be falsified if  a (nonlocal) hidden
variable  theory  were  able  to  explain  quantum  mechanics  in  the
future. As this seems unlikely, a  QRNG still seems the best bet for a
source of genuine and unique randomness.

We thank Prof.  N. Kumar, Prof. S. Banerjee and  Prof. V. Naryanan for
helpful discussions.

\end{document}